\documentclass[3p, times, authoryear]{elsarticle}
\graphicspath{ {./figures/} }
\usepackage{hyperref}
\usepackage{float}
\usepackage{amsmath}
\usepackage{amssymb}
\usepackage{verbatim} %comments
\usepackage{apalike}
\usepackage{mathrsfs,amsmath}
\restylefloat{figure}
\restylefloat{table}
\hypersetup{colorlinks}
\journal{Decision Support Systems}
\usepackage{color}
\usepackage{array}
\usepackage{algpseudocode}
\usepackage{algorithm}
\biboptions{authoryear}
\usepackage{setspace}

\begin{document}
\begin{frontmatter}

\title{NETpred: Network-based modeling and prediction of multiple connected \\ market indices}

\author[label1]{Alireza Jafari}
\ead{alireza.jafari7@ut.ac.ir}

\author[label1,cor1]{Saman Haratizadeh}
\ead{haratizadeh@ut.ac.ir}

\cortext[cor1]{Corresponding author.}
\address[label1]{Faculty of New Sciences and Technologies, University of Tehran, North Kargar Street, 1439957131 Tehran, Iran}

\begin{abstract}
%Prediction of the market index fluctuations has long been considered as a challenging task in the machine learning domain. Most of the traditional techniques use the historical data of the indices to learn models for the prediction of index fluctuations; however, in recent studies, it has been shown that modeling and analysis of stock-index relations can improve the prediction accuracy. One potential approach that has not been explored well in this line of research, is to generate and analyze graphical models of the stock-index relations. In this paper, we introduce NETpred, a novel framework that models the relations among a large set of stocks and indices as a heterogeneous network. To build this network, NETpred analyzes the historical fluctuation data and the current state of each pair of stocks to estimate how useful an edge can be in the subsequent information aggregation and label prediction process. It then assigns some initial labels to a set of carefully selected nodes and makes sure that the injected information is both accurate and appropriately distributed in the graph. A deep graph-based model is then trained over the partially labeled graph to forecast the next price movement of each stock as a label prediction process. Finally, the labels of the stocks are propagated to the index nodes to predict the index movement for all the index nodes in the graph.A comprehensive set of experiments on standard data sets of S\&P 500, NASDAQ, DJI, and NYSE show that NETpred significantly outperforms a diverse set of state-of-the-art baselines.
Market prediction plays a major role in supporting financial decisions. An emerging approach in this domain is to use graphical modeling and analysis to for prediction of next market index fluctuations. One important question in this domain is how to construct an appropriate graphical model of the data that can be effectively used by a semi-supervised GNN to predict index fluctuations. In this paper, we introduce a framework called NETpred that generates a novel heterogeneous graph representing multiple related indices and their stocks by using several stock-stock and stock-index relation measures. 
It then thoroughly selects a diverse set of representative nodes that cover different parts of the state space and whose price movements are accurately predictable. By assigning initial predicted labels to such a set of nodes, NETpred makes sure that the subsequent GCN model can be successfully trained using a semi-supervised learning process. The resulting model is then used to predict the stock labels which are finally aggregated to infer the labels for all the index nodes in the graph. 
Our comprehensive set of experiments shows that NETpred improves the performance of the state-of-the-art baselines by 3\%-5\% in terms of F-score measure on different well-known data sets.

\end{abstract}

\begin{keyword}
Index prediction \sep Deep learning
 \sep Graph neural network \sep Spectral clustering \sep Network-based prediction
\end{keyword}

\end{frontmatter}

\section{Introduction}
\label{introduction}
Financial decisions are among the most common yet serious decisions that people make in daily life. To make such decisions wisely, we usually gather information that can help us understand the current and future state of the target financial domain with which we are dealing. Stock markets are among the most important entities that can affect our financial decisions, so  market prediction models has been tuned into a major element in financial decision support systems. Stock market indices are among significant and informative financial factors and knowing the future direction of stock market indices helps traders to recognize and invest in a market with growth potential \citep{kia2020network}. Predicting the future behavior of market indices has been a challenging task in the machine learning domain.

A major class of modern and traditional index prediction models uses an index historical data, along with some technical indicators, to predict its future fluctuations \citep{hoseinzade2019cnnpred, NAM2019100, gunduz2017intraday}. In this approach, the stock-market relations are ignored and no effort is made to explicitly model and exploit such relationships for enhancing the index prediction performance. Surprisingly, this approach is in contrast with the fact that the price fluctuations of the stocks in an index are the basic elements that  define the behavior of that index.

In the past decade, some researchers have studied the effect of using historical data of important stocks in a market on the prediction performance of the market index \cite{zhong2017forecasting} \cite{hoseinzade2019cnnpred} \cite{lee2020effectively}. This approach relies on the designers' expertise in the selection of a set of representative stocks and the resulting model usually can not be generalized to consider an arbitrary set of stocks and indices.  

To tackle this problem some studies have tried to explicitly model the relations of stocks and indices using a graph structure and then analyze the resulting graphical model to make the desired predictions. In some pioneer works the common approach is to create a homogeneous network of several financial indices from the markets around the world and predict the future fluctuations of the indices using a label spreading/propagation technique over the resulting graph  \citep{kia2020network, tang2019global, kia2018hybrid, shin2013prediction}. Although these technique has achieved significant performance improvements over non-graphical forecasting methods, their simple prediction mechanisms fail to capture sophisticated patterns beyond what has already been modeled by the graph structure itself.

In recent years, introducing Graph Neural Networks has led to significant progress in the field of graph-based modeling and analysis. The power of GNNs in extracting useful prediction patterns from graph-structured data makes them ideal candidates to be applied for graph-based market prediction tasks. In a few recent studies, researchers have used GNNs for the prediction of market indices \citep{ matsunaga2019exploring, kim2019hats}. These models apply a supervised style of learning to train a GNN on simple graph structures representing connections of a single index with its sub-stocks. Although these models have achieved superior performance in prediction they still suffer from some flaws. First, the scope of the existing studies is limited to the prediction of a single index's movements and they can not directly model the relations of different indices that are naturally connected through their stocks. Second, they represent the data by simple and naive graph structures on which a GNN algorithm may not operate well. Finally, while some recent studies in the field of stock price prediction have introduced successful frameworks for semi-supervised GNN learning on graphs of stocks, it is not clear how such an approach can be taken in the field of index prediction, especially in a multi-index setting that requires larger graphical models.

To address these shortcomings we present NETpred, a novel graph-based index prediction framework whose main contributions can be summarized as follows:

%NETpred models the stock-stock and stock-index relations as novel heterogeneous graph structures. It then  clusters the stocks based on their  each experiencing a  injects some useful information to different spots analyzes the graph to applies a graph-based neural network to learn an index prediction model that aggregates information from relevant entities in the graph by analyzing the direct and indirect relations among sub-stocks with each other and with the indices. 

%The main contributions of the introduced framework, NETpred, can be summarized as follows:
\begin{itemize}
    \item{NETpred introduces a novel method to create an informative heterogeneous network that models the stock-stock and the index-stock relations in a single graph that implicitly captures the index-index indirect relations as well. The suggested method is not dependent to any specific set of stocks and can be easily adopted on any set of indices and stocks.}
    \item{To improve the performance of its semi-supervised learning process NETpred introduces a new initial label assignment approach that tries to evenly distribute the initial labels among the stocks that are experiencing different financial states.}
    \item{According to a comprehensive set of experiments on four famous and large indices, the overall prediction performance of the NETpred algorithm is remarkably higher than the state-of-the-art baselines.}

%    \item{NETpred introduces an innovative network-based method for predicting the stock market indices.}
%    \item{NETpred uses the impact of index sub-stocks as well as the indices on each other to predict the stock index and does not require index data.}
%    \item{NETpred in creating the network, in addition to the correlation in the direction of price movement between stocks, ensures the effectiveness of building new representations according to the network.}
%    \item{NETpred provides a new training method for the graph neural network that extracts more information from the graph by samples clustering.}
\end{itemize}

In the next section, we will review the related works. NETpred prediction algorithm and network construction method are presented in Section \ref{model}. Section \ref{experiment} provides detailed information about the datasets used in our experiments and experimental settings and presents and discusses the results.We conclude the article in Section \ref{conclusion}.

\section{Related Works}
\label{related_works}

%In this section, we are going to describe related literature work in the deep learning domain for index prediction. As we mentioned previously, many different machine learning techniques have been used for the prediction of stock market indices in the past years. 

Many different machine learning techniques have been used for the prediction of stock market indices. Most of these studies ignore the natural relation of the stocks in a market with the market index and the only source of information that they use for prediction is the historical data of the index itself \citep{NAM2019100, cao2019stock, KAO20131228, OLIVEIRA201662}. 

%Many different machine learning techniques have been used for the prediction of stock market indices. In most of these studies, regardless of the natural relation of the stocks in a market with the market index, the only source of information that is used for prediction is the historical data of the index itself. \cite{kara2011predicting} and \cite{arevalo2016high} used an artificial neural network trained by index historical data. \cite{moghaddam2016stock} tested ANNs with different structures to predict the market index, to show the superiority of deep artificial neural networks over shallow ones and they used index historical data. \cite{patel2015predicting} used ANN and two well-known classic Random Forest and SVM models to predict the movement of market indices. They showed that a useful mapping of index historical data and technical indicators could increase the accuracy of the forecast. 
% \cite{di2016artificial} applied ANN, LSTM, and CNN to predict the S\&P 500 index. They only considered the closing price as the input of their model, and the results showed that the use of convolution in the neural network is better in prediction than other deep models.
% \cite{cao2019stock} used CNN to predict the Hongkong Hang Seng Index (HSI). They have given index historical data to their models in lengths of 30 and 40 as a one-dimensional input.

% پاراگراف بالا فقط هیستوری شاخص
% پاراگراف پایین هستوری شاخص و سهام

As a step towards exploiting relevant market/stock information for index prediction, \cite{hoseinzade2019cnnpred} combined data from a set of indices and some stocks in a CNN-based framework. They predicted five famous U.S. stock market indices and showed that using a combination of related financial time series can improve the performance of index prediction significantly.
\cite{zhong2017forecasting} predicted the famous S\&P 500 index using ANN. They used PCA and two specific variations of it to generate better representations from the initial feature vectors. In addition to historical index data, they also exploited the return of eight large stocks of the S\&P 500 index as the input of their model and showed that exploiting information from the index's sub-stocks can enhance the performance of the index prediction.
 \cite{gunduz2017intraday} used CNN to predict the Istanbul Stock Exchange index. They considered a large set of technical indicators of indices and stocks and by introducing a clustering-based method for selecting a set of independent technical indicators as features. 
 \cite{lee2020effectively} trained the types of deep neural networks using stocks data from the S\&P 500 index and compared them with models trained using index data. Their experiments show that simpler neural networks trained with stocks data outperform neural networks trained on index data. However, like most of the above studies, they did not explicitly model the relationships between stocks and the index, and their method can not be directly applied to other sets of indices and stocks.
 
 %مدلهای گرافی و  مدلهای جی ان ان 
 
Graph-based forecasting methods are an emerging set of algorithms for predicting  market indices \citep{kia2020network, kia2018hybrid, shin2013prediction, skabar2013direction}. Basically, these methods model the relations of the financial entities as a graph structure and the problem of movement prediction as a node label prediction task on that graph \citep{zhu2005semi}. 
\citep{shin2013prediction} generated a network of some financial entities such as market indices and oil prices, in which each node is defined by a vector of technical indicators and each edge weight represents  the Euclidean distance between those vectors. They use a label-spreading technique over this graph to predict the fluctuations for each entity. \cite{park2013stock} took a similar approach to predicting the Korean Stock Exchange.
\cite{kia2018hybrid} used a correlation graph of a set of market indices, in which some nodes are labeled by their corresponding index's movement in the past hours and the labels of the rest of the nodes are predicted using a label propagation algorithm. In another study, \cite{kia2020network} a new graphical model of the index relations is introduced in which the nodes represent indices of the world markets and the edge weights are defined using an association rule mining technique. They predicted the unknown labels in the network with a variation of the PageRank algorithm.

Although this class of works represents the relationships of the financial entities as a graph structure, their label prediction process is simple and highly dependent on the structure of their underlying graphs that limits their pattern extraction power.
To address this issue, in a more recent category of studies graph neural networks (GNNs) have been applied to analyze the graphical models of financial data. GNNs are able to use the graph structure to generate more informative embeddings for a prediction task \citep{jafari2022gcnet}. However, the power of these models in forecasting the future behavior of financial time series has not been studied well yet, especially in the field of index movement prediction. 

\cite{chen2018incorporating} created a graph of company nodes based on financial investment information using which they trained a Graph Convolutional Neural Network model. The graph-based approach used in this algorithm is designed to predict the stock prices and is not directly applicable to predict the movement of market index values.
\cite{matsunaga2019exploring} predicted the Japanese Nikkei 225 market index using simple GNN structures. This work focuses on graphical modeling of the fundamental features in a market such as suppliers and customers and can not be applied where the only available  source of information is the historical data of the stock prices and index values.
\cite{kim2019hats} used a graph attention network to forecast stocks' movement and also tried to predict market indices by a graph pooling layer. For the infrastructure network, they used a classic graph of Wikidata in previous work \citep{vrandevcic2014wikidata}. 
%These models

Although these models, as the first GNN models used for stock market forecasting have achieved significant success, they have some limitations too: None of the reported research can model relations of multiple indices, use simple graph structures on which a GNN learning algorithm may not necessarily work well and can not support the semi-supervised style of learning. In the next section, we will introduce NETpred, which addresses these shortcomings by introducing an innovative approach to model the stock-index relationships as a heterogeneous graph of several stocks and indices, as well as a novel method for distributing some initial label information throughout the graph to improve the performance of the subsequent semi-supervised learning process.

%\section{Preliminaries}
%\label{preliminaries}

\section{Model description}
\label{model}

NETpred first creates a heterogeneous graph representing the relationships among several indices and their corresponding important stocks. Each node in this graph is assigned a feature vector summarizing its recent historical information. NETpred then  selects an appropriate group of the nodes to assign initial labels and then uses this partially labeled graph to train the final label prediction model. The resulting model is then exploited to label the stock nodes from which the labels of the index nodes are inferred. In the next subsections, we will explain each step in detail.

\subsection{Prediction Network Construction}

NETpred creates a heterogeneous graph that consists of index and stock nodes and two types of stock-index and stock-stock edges. We will refer to this graph as the "prediction network", as it is going to be used by the following steps of the algorithm in a label prediction process that predicts the price fluctuations of the stocks as well as the movements of the indices.
First, nodes are added to the graph for the target indices that exist in the dataset. Then, for each index, a set of stocks with the highest impact on the index value are added to the network, while each stock node is connected to its corresponding index nodes by an edge whose weight represents the importance of that stock's role in calculating the value of the index.
%, so that if the index uses the weighting method of market capitalization, stocks with the highest market capitalization will have the greatest impact in predicting the direction of the index. 
Each pair of stocks in the graph may also be connected by an edge whose weight represents a combination of their price correlation and a so-called influence factor.
%While Correlation is a value that tries to demonstrate stocks that move in the same direction. Furthermore, Influence is a value that tries to extract stocks whose data are useful for forecasting each other. The combination of these two values is used to construct the weights of the edges between stocks. 
Finally, a feature vector is assigned to each stock node, representing the last day status of the stock by a small set of technical indicators.

Here we denote the network as $N = (V , E)$, in which $V$ is the set of stock and index nodes while $E$ is the set of graph edges, including the stock-index edges $e_{I_i,S_j}$ and stock-stock edges $e_{S_i,S_j}$.

\subsubsection{Nodes}

In the network construction procedure, first, indices are added to the network. If there are  $n$ indices in the dataset, we have :  $indices = \lbrace I_{1}, I_{2}, ..., I_{n} \rbrace$. Each index is mapped to one node in the network. After that, the stock nodes are added to the network. For each index node, one hundred stocks with the highest impact in calculating the value of the index are added to the graph. Again, if there are  $m$ stocks in the dataset, we have :  $Stocks = \lbrace S_{1}, S_{2}, ..., S_{m} \rbrace$. So, the set of nodes is equal to: $V = \lbrace v_{I_{1}}, v_{I_{2}}, ..., v_{I_{n}},v_{S_{1}}, v_{S_{2}}, ..., v_{S_{m}} \rbrace$. 

%Then, we add the edges to the network. There are two types of edges in our network, stock-stock edges and stock-index edges. 
%In this work, we put one hundred stocks with the largest market cap in our data set from each index. We must note that the repetitive stock is placed once.

\subsubsection{Stock-Index edges}

Each index node is connected to its sub-stock nodes in the graph. To assign a weight to a stock-index edge, the effect of that stock on the value of the corresponding index is considered. An index aims to represent the overall market or a hypothetical portfolio of specific stocks. The value of the index is calculated from the prices of the underlying stocks, so changes in the prices of the stocks in the index determine the change in the index's value. Each index has its own calculation methodology, which is defined by the index provider. A capitalization-weighted index uses a company's market capitalization to determine how much impact a stock can have on the overall index results. In the composition of a capitalization-weighted index, large movements in the price of stocks for the largest companies can significantly impact the value of the overall index. In a price-weighted index, stocks with higher prices receive a greater weight in the index, regardless of the issuing company's actual size. The weights of the stock-index edges in the graph are calculated according to the calculation method defined by the corresponding index.

According to these definitions, we use equations (\ref{weight_MC}) and (\ref{weight_price}) to calculate the edge weights for the indices with capitalization weighting and price weighting respectively:

\begin{align}
\label{weight_MC}
weight_{I_{i},S_{j}}= \frac{log(\text{MC}_{S_{j}})}
{\sum_{\text{ all of stocks}}(log(\text{MC}))}
\end{align}
\begin{align*}
MC_{S_{j}}  \text{is the company market capitalization for the stock} j
\end{align*}

\begin{align}
\label{weight_price}
weight_{I_{i},S_{j}}= \frac{\text{price}_{S_{j}}}
{\sum_{\text{ all of stocks}}(\text{price})}
\end{align}

More detailed information about these two types of indices are  given in \ref{appendix:Index}.

\subsubsection{Stock-Stock edges}

The second type of edge in the graphical model of data is the stock-stock edge. To add these edges and define the corresponding edge weights, NETpred combines two relevance measures of Correlation and Influence ( as defined in \citep{jafari2022gcnet}).
%To demonstration the relationship between the directions of stocks movement, we use Pearson cross-correlation on the historical data of each stock pair. Also, to quantify the usefulness of a stock's data to predict another target stock, we define the so called Influence factor on each stock pair.
While a significant correlation between two stocks' historical data shows that their direction of price movement is usually correlated, a high influence value between two stocks shows that information from the recent price history of one of them can be helpful to predict the other one's next price movement.

if $Y_{s_i}$ represents the direction of daily price movements of stock $s_i$ in historical data the Pearson correlation between each pair of stocks $s_i$ and $s_j$ is defined by equation (\ref{equ_corr}):

\begin{align}
\label{equ_corr}
Correlation_{s_i,s_j}= \frac{\sum\limits_{k=1}^{d}
(Y_{s_i}^k-\bar{Y_{s_i}})(Y_{s_j}^k-\bar{Y_{s_j}})}
{\sqrt{\sum\limits_{k=1}^{d}
{(Y_{s_i}^k-\bar{Y_{s_i}})^{2}}
\sum\limits_{k=1}^{d}
{(Y_{s_j}^k-\bar{Y_{s_j}})^{2}}}}
\end{align}

The influence value for each pair of stocks, $s_i$ and $s_j$, is defined as an estimation of the average increase in prediction accuracy for each stock, that can be achieved by combining information from both stocks compared to when each stock's movement is predicted using its own historical information \citep{jafari2022gcnet}. 

To make such an estimation, for each stock a separate SVM model is trained on that stock's historical data (Table 1 shows the features used). Suppose that the resulting prediction accuracy values on validation data for stocks $s_i$ and $s_j$ are equal to $Acc^{raw}_i$ and $Acc^{raw}_j$, respectively.  In the next step, the two stocks feature vectors are averaged and used to train a new model for each of them. Suppose the prediction accuracy values in this setting are equal to $Acc^{Processed}_i$ and $Acc^{Processed}_j$. Then the influence value is defined by equation (\ref{equ_influ}):

\begin{align}
\label{equ_influ}
influence_{i,j} = \frac{1}{2} \sum_{k \in \{i,j\}} Acc^{Processed}_k - Acc^{raw}_k 
\end{align}
 
For each pair of nodes $v_{s_i}$ and $v_{s_j}$ the weight of the edge connecting them, $w_{i,j}$ is set equal to a weighted sum of their corresponding Pearson correlation and Influence values, as shown in equation \eqref{weight}:

\begin{align}
\label{weight}
weight_{s_{i},s_{j}} = weight_{s_{j},s_{i}}= ((\lambda*(influence_{s_i,s_j})) + ((1-\lambda)*(Correlation_{s_i,s_j}))
\end{align}

Where $0 < \lambda < 1$ is a coefficient that determines the significance of correlation vs. influence. 
%Figure 3 shows how different  $\lambda$ values can affect the resulting network. 
Since this graph is going to be ultimately processed by GCN, the intuition behind the weight calculation technique used in this step comes from the behavior of the aggregation phase in GCN, during which information from neighboring nodes is aggregated to be used as a feature vector for predicting the label of the target node. While the correlation value reflects how similar the behavior of the two stocks is, the influence value estimates how useful information from a neighbor is expected to be in the improvement of that prediction's accuracy. So we use a combination of these two factors as an edge weight based on which the amount of information flow among each pair of nodes is defined.

Following the approach used in \citep{jafari2022gcnet} to avoid noisy low-weighted connections, when all the edge weights are calculated, , NETpred removes the edges in ascending order of edge weights until removing the next edge will make the graph disconnected.  

\begin{figure*}[h]
\includegraphics[width=1\linewidth]{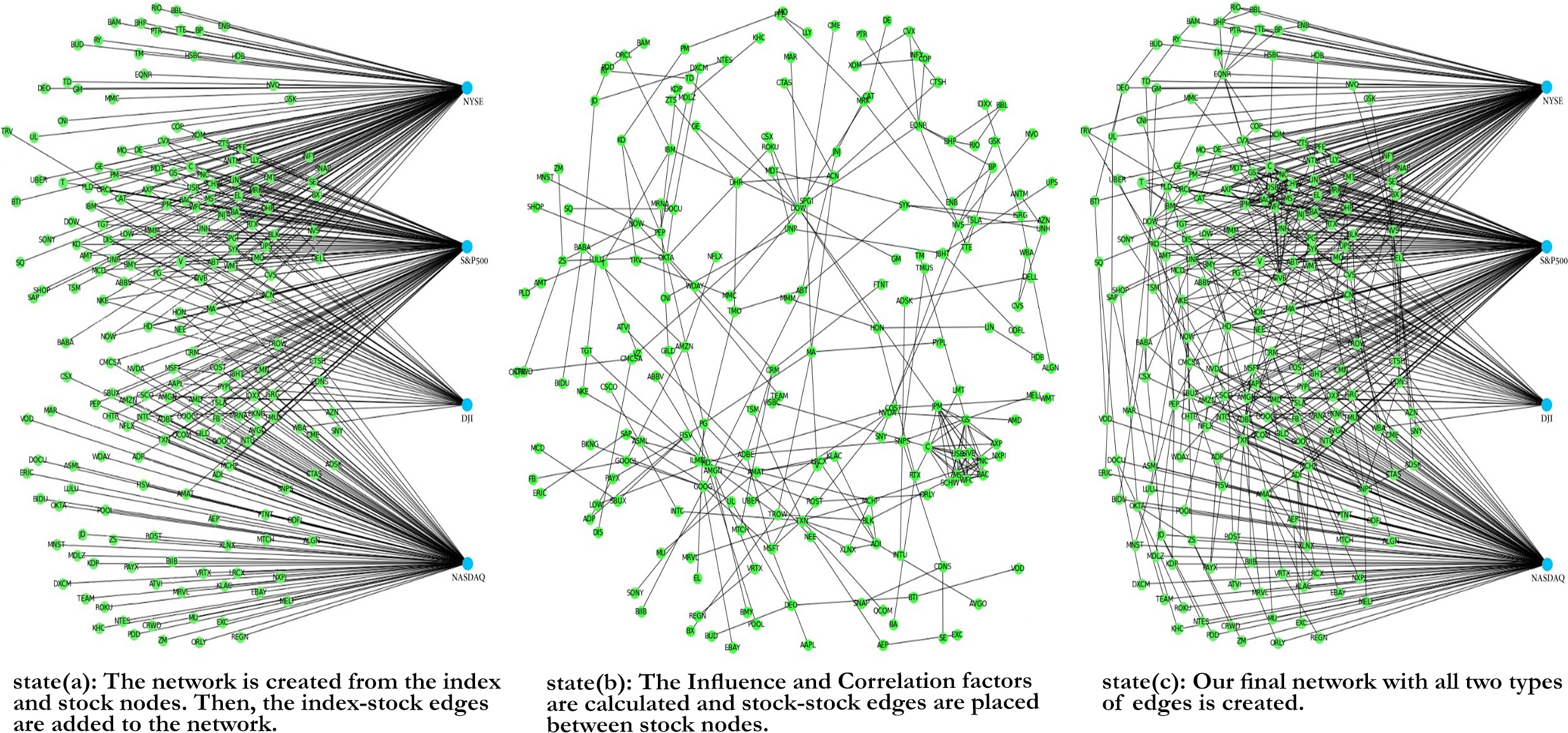}
\caption{Steps of network construction in NETpred model}
\label{graph_str}
\end{figure*}

The process of edge weight calculation can be repeated in certain time intervals to keep the graph up to date (in this study we update the weights every sixty days). Figure \ref{graph_str} shows the steps of creating the prediction network.

\subsection{Node features}

The state of each node in the network is represented by an initial feature vector containing some technical indicators and the stock’s last day price information. The complete set of node features are summarized in \ref{appendix:Input}.

\subsection{Initial information injection}
\label{cl}

Unlike most of the existing graph based methods, NETpred uses a semi-supervised GNN for label prediction mechanism on the generated prediction network. So it needs some initial labels for a subset of the nodes exist that can be used to train the label prediction model. Since the true labels are unknown for the target day, those initial labels need to be predicted by NETpred. Generally, any subset of nodes can be chosen for this initial information injection step, however since they can play a critical role in the overall success of the final model, NETpred selects them thoroughly. 

The node selection process is performed with two concerns in mind: First, to make sure that the initially assigned labels are as accurate as possible,  nodes whose next price movements can be predicted more accurately are preferred. Second, to cover a wider range of the state space in the training data, a diverse set of nodes must be selected from different financial states that the stocks are currently experiencing.

To address these two concerns, NETpred first clusters the stocks based on their current states(represented by the feature vectors of their corresponding nodes) and then selects the most accurately0 predictable stock node from each cluster to receive a label. In the following subsections, we will explain the clustering, node selection, and initial labeling steps in detail.

\begin{algorithm}[H]
\hspace*{\algorithmicindent}
\textbf {Input :}\\
\hspace*{\algorithmicindent}\hspace*{\algorithmicindent}\textbf{Dataset \& Unlabeled Graph G(V,E)} \\ 
\hspace*{\algorithmicindent}\textbf{Output :}\\
\hspace*{\algorithmicindent}\hspace*{\algorithmicindent}\textbf{Partially Labeled Graph G(V,E)}
\vspace{-7mm}
\\

\hrulefill
\begin{algorithmic}[1]
\State $D \gets$ {The set of distances between the feature vectors of each pair of stocks}
\State $SN \gets$ The KNN graph based on stock distances in D
\State $C \gets$ Run Spectral Clustering on SN \Comment{C contains the set of nodes clusters}

\For{each Cluster $C_i$ in C}
\For{each stock $s_i$ in $C_i$}
%\State $score_{s_i}\gets RF\_Predictor.fit(train\_data_{s_i}).score(validation\_data_{s_i})$ 
\State $score_{s_i} \gets$ The movement prediction accuracy on validation data %\Comment{using Random Forest model}
\EndFor
\State $s_{best}\gets$ the stock $s_i$ in $C_i$ with the highest $score_{s_i}$
\State $initial\_label(s_{best})\gets$ Predict the next movement of $s_{best}$  \Comment{add the node label to G}
\EndFor

\State $return (G(V,E))$ \Comment{Partially labeled graph}
\caption{Initial\ Information\ Injection\ algorithm}
\end{algorithmic}
\label{alg:GCNET}
\end{algorithm}

\subsubsection{Clustering the stocks}
\label{sc}

NETPred tries to identify groups of stocks that are experiencing similar financial states. One good candidate for this goal is to use the spectral clustering method which is able to cluster the nodes of a graph. Although it is possible to use the prediction network structure that we have built in the previous steps of the algorithm as the input of the spectral clustering algorithm, it is clearly not the best choice. The prediction network represents the relations of the stocks based on two factors of price correlation and the influence value which do not necessarily reflect the similarity of the current states of the stocks in the market.  So NETPred builds another graphical representation of data that models the similarity of stocks' states and use the spectral clustering method to cluster the stock nodes in this second graph structure. Since this graph is built based on the similarity of the current states of the stock, here we refer to it as the similarity network.

Suppose our data contains $n$ stocks. The similarity of each pair $s_i$ and $s_j$ of stocks is defined by some distance measure (here we use radial basis functions). Then the similarity network $SN$ is constructed as a $k$-nearest neighbor graph in which each stock node is connected to its $k$ most similar stock nodes. The weight of an edge in this graph is defined by the similarity of the feature vectors of the nodes it connects.  Generally, the algorithm is not too sensitive to the choice of $k$, and a small number like 5 or 10 usually works well. 

The resulting similarity network is then partitioned by the spectral clustering method to find the clusters of stocks. To do so, the Laplacian matrix of SN is defined. Then the first $k$ eigenvalues of this Laplacian matrix are calculated and stacked to represent the set of nodes in a new feature space. These new node representations are then used by a traditional clustering algorithm to cluster the nodes into $k$ clusters.

%as: 
% \begin{align}
% L_{Sym} = D^{-1/2} L D^{-1/2} = I - D^{-1/2} A D^{-1/2}
% \end{align}

% where $D$ is the degree matrix and $A$ is the adjacency matrix of the graph SN. 

It is important to note that While the prediction network is designed to provide useful paths for later information propagation steps in the algorithms, the similarity network is applied to model the similarities among the current states of the stocks, which is essential for our intended goal of clustering that is to identify groups of stocks that are currently experiencing similar states according to their recent prices.

\subsubsection{Initial labeling}
\label{lebel}

Suppose NETpred has extracted $k$ clusters of stocks from the similarity network. For each cluster, it trains a set of random forest models each predicting one of the stocks in the cluser and finding the stock that is predicted by its corresponding model most accurately, over 20 days of validation data. It then predicts the next price movement for that most predictable stock in each cluster using its random forest model and uses the resulting predicted direction of movement as the initial label assigned to that stock, The output of this step is the partially labeled prediction network in which $k$ nodes are assigned initial labels while the rest of the nodes are left unlabeled.
%By doing so, the most predictable stocks in each cluster are labeled with the most differences from the feature vector $X$ perspective. 
This process is repeated for each day of prediction, and on different days, different stocks may be selected and assigned initial labels. 

\subsection{Label prediction for stock nodes}

In the next step, we use the GCN algorithm as a semi-supervised framework for representation learning on the partially labeled prediction network to predict the final labels for all the stocks. The input of the model contains a graph with a number of labeled nodes and initial feature vectors $X$.  Each layer of GCN integrates information available in the neighborhood of node to obtain a new representation for that node, that is used to train a label prediction model. The resulting trained model is then applied to predict the labels for all the stock nodes in the network. 

The general architecture of GCN consists of a series of convolutional layers, each followed by Rectified Linear Unit (ReLU) activation functions. In this paper, we apply a two-layer GCN model for stock price movement prediction that can be defined as:

\begin{align}
\label{eq:GCN_model}
Y = softmax(\widehat{A} ReLU(\widehat{A}XW^{0})W^{1})
\end{align}

where $\widehat{A} = \tilde{D}^{-\frac{1}{2}} \tilde{A} \tilde{D}^{-\frac{1}{2}}$. Here, $\tilde{A} = A +I_{N}$, in which $\tilde{A}$ is the adjacency matrix of the graph $G$,  with added self-connections, $I_{N}$ is the identity matrix, $\tilde{D} = \sum_{j} \tilde{A}_{ij} $. 
$W^{0} \in \mathbb{R}^{|X|\times C_{0}}$ is an input-to-hidden weight matrix for a hidden layer with $C_{0}$ feature maps and $W^{1} \in \mathbb{R}^{C_{0}\times C_{1}}$ is a hidden-to-output weight matrix. The value of $C_{1}$ for prediction of the two classes (rise and fall) is 2.

We train the network using cross-entropy loss over predictions for all stocks modeled in the network:
\begin{align}
\label{eq:LOSS}
Loss = - \sum_{i \in Z_{L}} \sum^{F}_{f=1} Z_{if} lnY_{if}
\end{align}

where $Z_{L}$ is the set of stocks with initial labels, $F$ is the classes of stock price movement direction.
Considering the number of features and training examples, we used 4 channels for the first layer and 2 channels to predict 2 classes in the second layer.
%The activation function for the first layer is RELU and for the second layer is Sigmoid.

\subsection{Index movement prediction}

\begin{figure*}[h]
\centering
\includegraphics[width=1\linewidth]{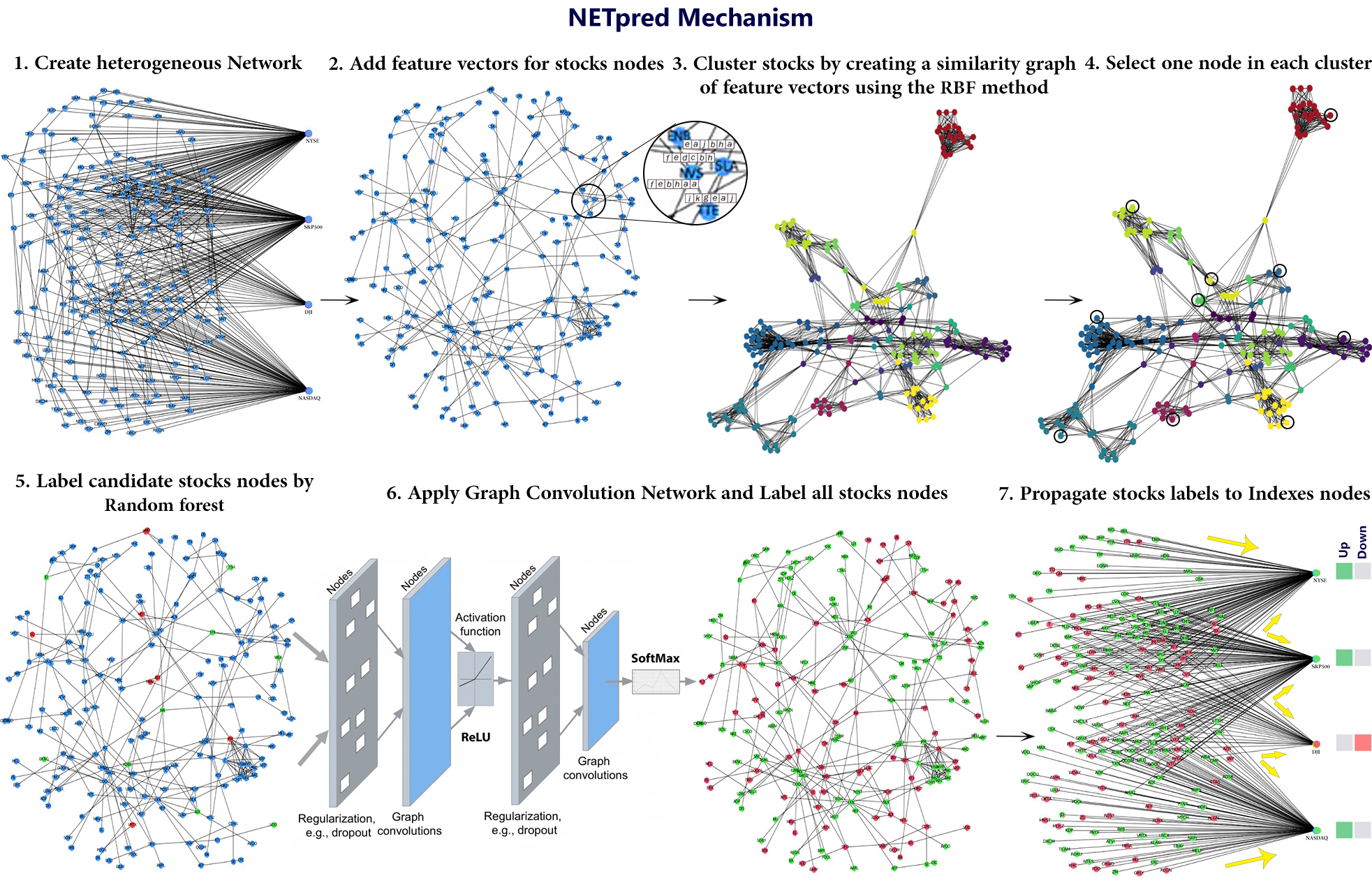}
\caption{Architecture of the components of the NETpred model}
\label{schema}
\end{figure*}

It is possible to treat the index nodes like the stock nodes and predict the labels of the index nodes using the trained GCN model. However since the indices are entities different entities from the stocks, and they also have different structural attributes in the graph, patterns extracted by GCN to predict the labels of the stock nodes may not generalize well to index movement prediction. So NETPred predicts the final label of each index node by propagating the predicted stock labels to the index nodes through the stock-index edges, as defined by the equation (\ref{index_label}):

\begin{align}
\label{index_label}
Label_{Index_{i}}= Sign(\sum\limits_{k \in N_i} w_{ik} L_{s_k})
\end{align}

where $N_i$ is the set of the index node's neighbors, which are the nodes representing the stocks of the index. $w_{ik}$ is weight of the edge connecting nodes $i$ and $k$ and $L_{s_k}$ is the predicted label of the stock node $k$. Please note that while labels of each index's stocks directly affect the label of that index, the stocks of the other indices can have an indirect effect through the stock-stock relationships defined in the graph. 
As mentioned earlier the consecutive steps of clustering and Initial Labeling, stock label prediction, and index label prediction are repeated for each prediction day.
Figure \ref{schema} shows the component architecture and steps of NETpred.

\section{Experiments}
\label{experiment}

In this section, we explain the dataset, the evaluation process, and the results of the NETpred model.

\subsection{Data gathering and preparation}
\label{data}

In our experiments, we used four well-known market indices which include S\&P 500 index, NASDAQ Composite, Dow Jones Industrial Average, and NYSE Composite. NETpard does not need explicit historical data for the indices, instead, for each index, it exploits the historical data for 100 sub-stocks with the largest impact on the value of that index (duplicate stocks are included in the data set only once and are represented by a single node in the graphical models of the data), that form 200 different stocks, whose historical price data has been obtained from Yahoo Finance.  
Our proposed model runs separately for each test day in a testing period from 01/01/2021 to 30/06/2021. For training the random forest models, NETpred uses the data from 01/01/2017 to 30/09/2020 to create the graph (that is rebuilt again every 30 days). Also, it uses the data from 15/09/2020 to 29/09/2020 for validation and to set the hyperparameters. %Our dataset is available at the following address\footnote{\href{https://github.com/alireza-jafari/NETpred-Dataset}{https://github.com/alireza-jafari/NETpred-Dataset}}.

\subsection{Implementation and Parameters of NETpred}
We apply Adam optimizer to train the GCN and to avoid over-fitting we applied L2 regularization and dropout techniques. The weights of the GCN have been initialized by Glorot uniform initializer \citep{glorot2010understanding}. For GCN implementations we used Keras and Spektral \citep{grattarola2020graph} libraries.
 The number of top \%n nodes that are labeled by the random forest models is set using the validation data. Other parameters like learning rate and dropout rate have been set to widely-used values in the literature.

\subsection{Evaluation Metrics}
Although Accuracy is the most common evaluation metric used in the domain of classification, in an imbalanced dataset, it may be biased toward the models that tend to predict the more frequent class. So, following previous work in this field \citep{gunduz2017intraday, hoseinzade2019cnnpred}, to evaluate the performance of the algorithms, we use the Macro Averaged F-Measure that is the mean of F-measures calculated for each of the two classes.

\subsection{Baselines}
We compare our model with a diverse set of modern baseline algorithms, including network-based models, deep learning models, artificial neural network models, and models that use feature selection:
\begin{itemize} 
\item \textbf{PCA+ANN:} The first baseline algorithm is the one reported in \citep{zhong2017forecasting}. In this algorithm, the initial data is mapped to a new feature space using PCA. PCA is a Linear dimensionality reduction algorithm that uses singular value decomposition of the data can project it to a lower-dimensional space.  Then, the PCA+ANN with using the generated representation of data trains a shallow ANN for making predictions.

\item \textbf{CNN-cor:} The CNN-cor algorithm is a deep learning algorithm with two-dimensional input \citep{gunduz2017intraday}.In the first step, CNN-cor clusters all features into different groups and rearranges similar features in input according to the values of the feature correlations. The resulting representation of the data is then used by a CNN with a deep structure for prediction.

\item \textbf{HATS:} hierarchical graph attention method is a deep learning algorithm that uses a relational modeling module with initialized node representations. Their model selectively aggregates information on different relation types and finally uses a pooling layer for index prediction \citep{kim2019hats}.

\item \textbf{CNN-pred:}The CNNpred is a deep model that uses a diverse set of financial variables, including technical indicators, stock indices, futures contracts, etc \citep{hoseinzade2019cnnpred}. CNNpred constructs three and two-dimensional input tensors. Then, the input tensors are fed into a specified CNN model to make predictions. They use a combination of related time series data as input to their model to consider the relationship between stocks.

\item \textbf{exLift+DiMexRank:} exLift + DiMexRank is a network-based model for the prediction of market indices. The exLift network encrypts relationships between markets' indices using Association rule learning and predicts the direction of market index movement using PageRank algorithm \citep{kia2020network}.  The initial labeling step of the algorithm is designed based on the geographical and temporal distance of different countries.

\item \textbf{LSTM-forest:} A multi-task model that simultaneously predicts stock market return values and movement directions to improve the predictability. Similar to the structure of the random forest model, they train a large number of LSTM models and generate the final direction of the index by calculating the average of the output of each model. For each LSTM model, a small random set of features prepared for prediction is selected \citep{park2022stock}.

\end{itemize}

\subsection{Results and discussion}

Table \ref{tab:RESULT} shows the performance of the NETpred and baseline methods on S\&P 500 index, NASDAQ Composite, Dow Jones Industrial Average and NYSE Composite datasets averaged on different runs. LSTM-forest and 3D-CNNpred that are popular baseline deep models achieve higher prediction performances than other baselines. LSTM-forest applies a hundred separate LSTM models as well as a large set of features to predict the direction of an index's movement. Similarly, CNNPred and CNN-cor are deep CNN-based models receiving a diverse set of input information including various indicators to generate high-level embeddings and make predictions. The superiority of our model over these sophisticated models as well as other non-graphical baselines shows the power of the suggested graph-based modeling in combining simple pieces of information to infer useful embeddings and accurate predictions. 

As the result table shows, DimexRank also achieves a good place in the ranking of the algorithms, after a couple of deep models, as a graph-based prediction algorithm with a simpler graph analysis module compared to NETpred. The satisfactory performance of DimexRank, affirms the usefulness of the graphical modeling in this domain, even if a simple label prediction mechanism is applied on the underlying graph. Also, the superiority of our model over DiMexRank shows the positive effect of using the suggested network structure and the graph analysis technique in NETpred.  As we mentioned before, discovering and combining latent information from different stocks, whose relations are modeled as a graph structure, is the core ability of GCN that we exploited in our suggested algorithm, and as the experimental observations show, it can lead to significant improvement of the prediction performance.

\setlength{\tabcolsep}{10pt}

\begin{table*}[h]
\centering
\caption{Comparison of NETpred results based on F-measure with baselines}
\label{tab:RESULT}
\begin{tabular}{lcccc}
\hline
Model & NYSE & NASDAQ & S\&P500 & DJI
\rule{0pt}{4mm}%
\\[1.5mm]
\hline
\hline

PCA+ANN \citep{zhong2017forecasting} & 0.4359  & 0.4281  &  0.4332 & 0.4185
\\
CNN-cor \citep{gunduz2017intraday} & 0.4413  & 0.4227  &  0.4392 & 0.4421
\\
HATS \citep{kim2019hats} & 0.4196  & 0.4318  &  0.4444 & 0.4512
\\
MixDiMex \citep{kia2020network} &  0.4665 & 0.4215 & 0.4491 & 0.4558
\\ 
2D-CNNpred \citep{hoseinzade2019cnnpred} & 0.4609  & 0.4707  &  0.4673 & 0.4592
\\ 
3D-CNNpred \citep{hoseinzade2019cnnpred} & 0.4733  & 0.4654  &  0.4724 & 0.4839
\\
LSTM-forest \citep{park2022stock} & 0.4701  & 0.4682  &  0.4934 & 0.4899
\\
\hline

\textbf{NETpred} (this work) & \textbf{0.5255} & \textbf{0.5179} & \textbf{0.5228} & \textbf{0.5267}
\\
\hline
\end{tabular}
\end{table*}

\subsubsection{Network analysis  }

\begin{figure*}
\centering
\includegraphics[width=0.7\linewidth]{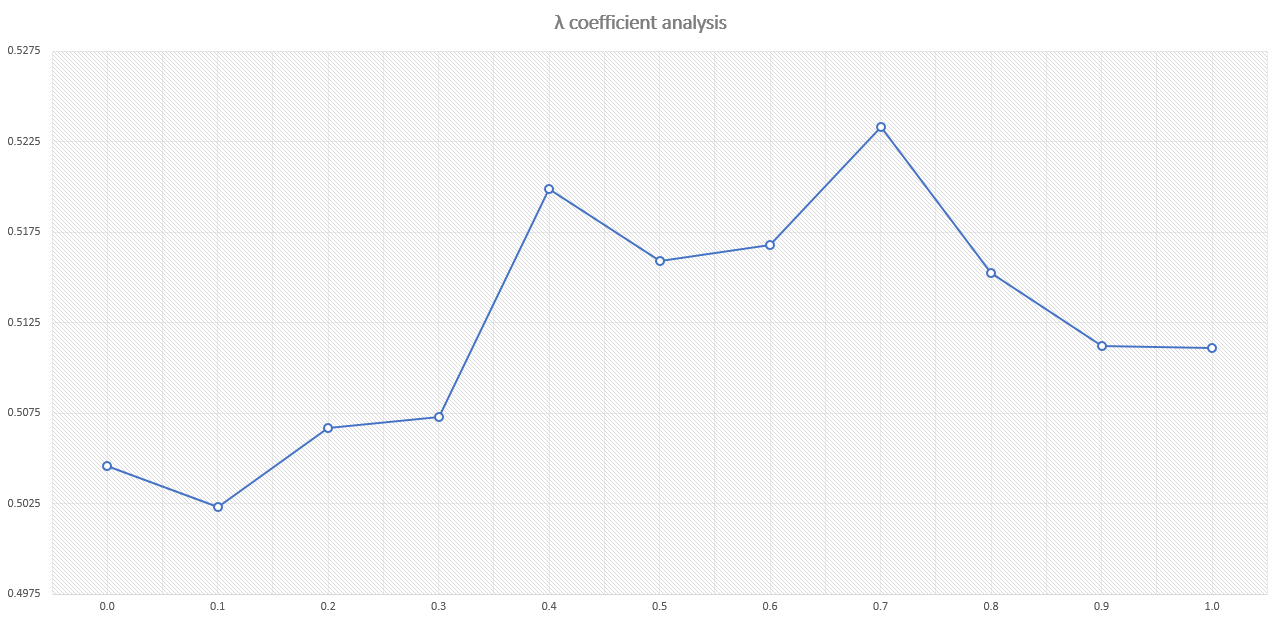}
\caption{NETpred model's f-measure chart in terms of different $\lambda$ values. Comparison of different $\lambda$ values in creating the network in NETpred algorithm}
\label{lambda_values}
\end{figure*}

To demonstrate the effect of integrating correlation and influence factors, we performed several experiments using different $\lambda$ values between 0 and 1. A $\lambda = 0$ means the network is constructed using the correlation factor while a $\lambda = 1$ means it is built using the influence factor. 
%We have compared the results with LSTM-forest as the best baseline that has an average f-measures of 0.4804. 
The average performance on 4 indices used in this work in terms of different $\lambda$ values has been illustrated in figure \ref{lambda_values}. It shows the highest f-measure belongs to the middle of the curve, where  the $\lambda$ is 0.7 and both criteria play their roles in generating the network. 

\subsubsection{prediction method analysis}

To demonstrate the performance of the index label prediction method we tested two other procedures. 
In the first approach, we treated the index nodes in the graph just like the stock nodes, and used the same 3-layered GCN model that is trained for prediction of stock price movements to also predict the index movements.

In the second approach, we used a graph pooling layer after two GCN layers to aggregate the representation of stock feature vectors in order to generate embeddings for the state of the index nodes. To construct the representation using the graph pooling
layer for each index, we considered only its sub-stocks, which divides the prediction network into four separate sub-
graphs.
%To construct the representation using the graph pooling layer for each index, we considered only its sub-stocks, which divides the prediction network into four separate sub-graphs and we should use separate models for the prediction of each index. 
Then, we used those index embeddings to  train models for prediction of the index movement direction over the training days data. Finally the resulting prediction models were used to forecast the test days' movements.

\setlength{\tabcolsep}{10pt}

\begin{table*}[h]
\centering
\caption{Comparison of NETpred prediction method results based on F-measure}
\label{tab:RESULT3}
\begin{tabular}{lcccc}
\hline
Model & NYSE & NASDAQ & S\&P500 & DJI
\rule{0pt}{4mm}%
\\[1.5mm]
\hline
\hline
GCN + indices as ordinary nodes & 0.4679  & 0.4657  &  0.4625 & 0.4723
\\
GCN + a graph pooling layer & 0.4814  & 0.4717  &  0.4792 & 0.4981
\\
\textbf{NETpred} (this work) & \textbf{0.5255} & \textbf{0.5179} & \textbf{0.5228} & \textbf{0.5267}
\\
\hline
\end{tabular}
\end{table*}

As shown in the table \ref{tab:RESULT3}, treating the index nodes just as other stock nodes in the graph decreases the accuracy of prediction. This shows that the prediction patterns extracted by GCN for stock label prediction can not be directly generalized to the index movement prediction. The reason could be that the index nodes differ from the stock nodes in several ways: the index nodes are connected to many other nodes in the graph, they are naturally different from their neighboring stock nodes and they don't have initial labels in the training data and so the GCN trained model is actually not "prepared" for index node label prediction. The results also show that even embedding the index nodes and training models to predict their labels, can not improve the performance of the algorithm. The reason is probably that the original neighbor-based configuration of NETpred for deriving the index nodes labels a more agile method than the alternative model based technique that can suffer from over-generalization.

\subsubsection{Node selection method analysis}

In our model, we use Spectral clustering on a similarity-based graph of stocks to categorize the nodes and apply a Random forest algorithm to find the most predictable node in each cluster for initial label assignment. To evaluate the efficiency of this node selection procedure, we have tested three other alternative approaches: random selection of the nodes, node selection by spectral clustering of the prediction network (instead of the similarity based graph) of stocks or just selecting the most predictable stocks without clustering the graph nodes.

\setlength{\tabcolsep}{10pt}

\begin{table*}[h]
\centering
\caption{Comparison of NETpred node selection method results based on F-measure}
\label{tab:RESULT2}
\begin{tabular}{lcccc}
\hline
Model & NYSE & NASDAQ & S\&P500 & DJI
\rule{0pt}{4mm}%
\\[1.5mm]
\hline
\hline
random node selection & 0.4754  & 0.4657  &  0.4853 & 0.4737
\\
most predictable node selection & 0.5012  & 0.4830  &  0.5045 & 0.4989
\\
node selection by clustering the prediction netwrok & 0.5123  & 0.5011  &  0.5096 & 0.5129
\\
\textbf{Original NETpred} & \textbf{0.5255} & \textbf{0.5179} & \textbf{0.5228} & \textbf{0.5267}
\\
\hline
\end{tabular}
\end{table*}

As summarized in the table \ref{tab:RESULT2}, random node selection as the simplest method drastically decreases the model's performance. While selecting the most predictable nodes and clustering the prediction network achieve acceptable performance, the original method is still significantly better than other variations of node selection, as we expected.

\section{Conclusion}
\label{conclusion}

In this paper, we presented NETpred, a novel network-based framework for the prediction of the daily movements of stock indices.  Our suggested framework uses the index-specific stock impact  measures, price correlations an the influence measure to generate a novel network structure for modeling the relations among the stock price movements and the index fluctuations for an arbitrary set of stocks and indices. To be able to use a semi-supervised learning algorithm for label prediction over this graph, NETpred injects some initial labels to the graph and uses a clustering approach to make sure that the initial labels are appropriately distributed among stocks experiencing different states. The partially labeled network is then analyzed by a semi-supervised GCN model to predict the next day's price movement for all the stock nodes. Finally NETpred propagates and aggregates  the stock labels through the weighted stock-index edges to infer the next direction of movement for each index. Our experiments showed that the suggested network-based algorithm can lead to better prediction of index fluctuations compared to the state-of-the-art baselines including modern deep models. The results also confirmed that the clustering-based information injection method as well as the final information propagation for index label prediction moth make a positive contribution in the NETpred success.

It can be subject of further research to generate more informative graph-based models of the stocks and index relations, possibly by analyzing the historical data along with other available information such as fundamental indicators or experts' market analysis reports. Also using more sophisticated network analysis techniques to improve the performance of initial information injection seems to be a promising line of research. As a more general issue, in this study, we constructed the prediction network independently from the related process of initial labeling. Since these two problems are naturally connected, we suggest to study how they can be solved together.

\bibliography{sample}

\appendix

\section{Market Index}
\label{appendix:Index}
The market index is a measurable and traceable number, which aims to represent the overall market or a hypothetical portfolio of specific stocks. A wide variety of investors use market indices for observing the behavior of the financial markets, using which they manage their trades. The value of the index is calculated from the prices of the underlying stocks. Each index has its own methodology of calculation, which is defined by the index provider. 

A capitalization-weighted index uses a company's market capitalization to determine how much impact that particular security can have on the overall index results. Market capitalization is derived from the value of outstanding shares. In the composition of a capitalization-weighted index, movements in the price of shares for the largest index companies can significantly impact the value of the overall index. To calculate the value of a capitalization-weighted index, we should first multiply each component's market price by its total outstanding shares to arrive at the total market value. The proportion of the stock's value to the overall total market value of the index components provides the weighting of the company in the index. Many stock market indices are capitalization-weighted indices, including the S\&P500 Index, the New York Stock Exchange Index, and the Nasdaq Index, which  we have used in our experiments. The capitalization of the stocks in the Nasdaq Index and New York Stock Exchange index is available on the Nasdaq website\footnote{\href{https://www.nasdaq.com/market-activity/stocks}{https://www.nasdaq.com/market-activity/stocks}}. Also, the subset stocks of the S\&P500 can be found on the Wikipedia website\footnote{\href{https://en.wikipedia.org/wiki/List_of_S&P_500_companies}{https://en.wikipedia.org/wiki/List\_of\_S\&P\_500\_companies}}.

In a price-weighted index, stocks with higher prices receive a greater weight in the index, regardless of the issuing company's actual size or the number of shares outstanding. To calculate the value of a simple price-weighted index, we should find the sum of the share prices of the individual companies, and divide it by the number of companies.  Accordingly, if one of the higher-priced stocks has a huge price increase, the index is more likely to increase even if the other stocks in the index decline in value at the same time. For example, a stock that increases from \$110 to \$120 will have the same effect on the index as a stock that increases from \$10 to \$20, even though the percentage move for the latter is far greater than that of the higher-priced stock. Higher-priced stocks exert a greater influence on the index's, or the basket's, overall direction. One of the most popular price-weighted stocks is the Dow Jones Industrial Average, which consists of 30 different stocks and we have considered it in our experiments. The weights of stocks in the Dow Jones Industrial index are available from Wikipedia\footnote{\href{https://en.wikipedia.org/wiki/Dow_Jones_Industrial_Average}{https://en.wikipedia.org/wiki/Dow\_Jones\_Industrial\_Average}}, although they can be easily calculated by stock prices.

\section{Input features}
\label{appendix:Input}

The complete set of features used for representing each node vector and training the models has been summarized:

%\begin{table}[h]
%\setlength{\extrarowheight}{2pt}
%\centering
%\caption{Technical indicators and price information used in various parts of this work}
%\label{tab:input}
%\small
%\begin{tabular}{llcc}
%\hline
%Indicator & Description & SVM \& RF & feature vector X
%\rule{0pt}{4mm}%
%\\[1.5mm]
%\hline

%OP & Open price  & $\bullet$ & 
%\\
%HP & High price & $\bullet$ &
%\\
%LP & Low price & $\bullet$ &
%\\
%CP & Close price & $\bullet$ &
%\\
%Volume & Trading volume & $\bullet$ &
%\\
%CCI & Commodity Channel Index & $\bullet$ &
%\\
%SAR & Stop and reverse index & $\bullet$ &
%\\
%ADX & Average directional movement  & $\bullet$ &
%\\
%MFI & Money flow index & $\bullet$ &
%\\ 
%RSI & Relative strength Index & $\bullet$ &
%\\
%SK & Slow stochastic \%K & $\bullet$ &
%\\
%SD & Slow stochastic \%D & $\bullet$ &
%\\
%RSI-S  & RSA indicator signal & $\bullet$ & $\bullet$
%\\
%BB-S  & Bollinger bands indicator signal & $\bullet$ & $\bullet$
%\\
%MACD-S  & MACD indicator signal & $\bullet$ & $\bullet$
%\\
%SAR-S  & SAR indicator signal & $\bullet$ & $\bullet$
%\\
%ADX-S  & ADX indicator signal & $\bullet$ & $\bullet$
%\\
%S-S  & Stochastic indicator signal & $\bullet$ & $\bullet$
%\\
%MFI-S  & MFI indicator signal & $\bullet$ & $\bullet$
%\\
%CCI-S  & CCI indicator signal & $\bullet$ & $\bullet$
%\\
%V-S  &Sign(Volume -Avg(last 5 days)) & $\bullet$ & $\bullet$
%\\
%CPOP-S  & Sign(CP-OP) & $\bullet$ & $\bullet$
%\\
%CPCPY-S  &  Sign(CP-Closing price yesterday) & $\bullet$ & $\bullet$
%\\
%\hline
%\end{tabular}
%\end{table}

\setlength{\tabcolsep}{2pt}
\begin{table}[h]
\setlength{\extrarowheight}{1pt}
\centering
\caption{Technical indicators and price information used in various parts of this work}
\label{tab:input}
\small
\begin{tabular}{llcc|llcc}
\hline
Indicator & Description & SVM \& RF &  GCN & Indicator & Description & SVM \& RF &  GCN
\rule{0pt}{4mm}%
\\[1.5mm]
\hline
\hline
OP & Open price  & $\bullet$ & & RSI-S  & RSA indicator signal & $\bullet$ & $\bullet$ 
\\
HP & High price & $\bullet$ & & SAR & Stop and reverse index & $\bullet$ & 
\\
LP & Low price & $\bullet$ & & SAR-S  & SAR indicator signal & $\bullet$ & $\bullet$
\\
CP & Close price & $\bullet$ & & SD & Slow stochastic \%D & $\bullet$ & 
\\
Volume & Trading volume & $\bullet$ & & SK & Slow stochastic \%K & $\bullet$ & 
\\
CCI & Commodity channel index & $\bullet$ & & BB-S  & Bollinger bands indicator signal & $\bullet$ & $\bullet$
\\
CCI-S  & CCI indicator signal & $\bullet$ & $\bullet$ & V-S  &Sign(Volume -Avg(last 5 days)) & $\bullet$ & $\bullet$
\\
ADX & Average directional movement & $\bullet$ & & S-S  & Stochastic indicator signal & $\bullet$ & $\bullet$
\\
ADX-S  & ADX indicator signal & $\bullet$ & $\bullet$ & CPOP-S  & Sign(CP-OP) & $\bullet$ & $\bullet$
\\
MFI & Money flow index & $\bullet$ & & MACD-S  & MACD indicator signal & $\bullet$ & $\bullet$
\\ 
MFI-S  & MFI indicator signal & $\bullet$ & $\bullet$ &  CPCPY-S  &  Sign(CP-Closing price yesterday) & $\bullet$ & $\bullet$
\\
RSI & Relative strength Index & $\bullet$ & &
\\

\hline
\end{tabular}
\end{table}

\end{document}